\documentclass[12pt]{article}
\usepackage{graphicx,amsmath,amssymb,amsfonts,amsthm,float,mathabx, array,accents,mathtools,wasysym,tipx,tipa,leftidx,tensor}
\usepackage[toc,page]{appendix}
\usepackage[usenames,dvipsnames]{color}                                
\usepackage[left= 1in,right=1in,top=1in,bottom=1in]{geometry}          
\numberwithin{equation}{section} 
\usepackage{hyperref,cite,cleveref}
\hypersetup{                 
	colorlinks=true,         
	linkcolor=blue,          
	citecolor=red,           
	urlcolor=Violet          
}
\theoremstyle{plain}

\theoremstyle{definition}

\newcommand{\dd}{\mathrm{d}}
\newcommand{\tho}{\text{\textthornvarii}}

\newcommand{\conj}[1]{\bar{#1}}
\newcommand{\edh}{\text{\dh}}
\begin{document}
\title{The Bach equations in Spin-Coefficient form}
\author{Hamish Forbes\\
		City, University of London,\\
		UK.\\
		\href{mailto:hamish.forbes@city.ac.uk}{hamish.forbes@city.ac.uk}}
\maketitle
\begin{abstract}
Conformal gravity theories are defined by field equations that determine only the conformal structure of the spacetime manifold.  The Bach equations represent an early example of such a theory, we present them here in component form in terms of spin- and boost-weighted spin-coefficients using the compacted spin-coefficient formalism.  These equations can be used as an efficient alternative to the standard tensor form.  As a simple application we solve the Bach equations for pp-wave and static spherically symmetric spacetimes.
\end{abstract}
\section{Introduction}
{In the hope to describe more varied phenomena, the theory of General Relativity has been generalised in many different ways, to name just a few, connections with torsion were considered, scalar-tensor theories were explored and higher-order curvature terms in the Lagrangian were included, see~\cite{capozziello_extended_2011,schmidt_fourth_2007} respectively for a modern review of extensions of General Relativity and a historical overview of fourth-order derivative gravity.
An example of the last was proposed by Bach~\cite{bach_zur_1921} in 1921 where he introduced a set of field equations now called the Bach equations, which can be derived from an action formed from the square of the Weyl tensor.  Due to the conformal invariance of the action, the Bach equations are also conformally invariant, hence Bach's theory is an early example of a conformal gravity theory.  More recently, several authors have found exact solutions to the Bach equations under particular conditions, for examples see~\cite{fiedler_exact_1980,mannheim_exact_1989,mannheim_solutions_1991}.\\The field equations can be solved using various different mathematical formalisms, e.g. the original coordinate approach to tensor calculus, its tetrad based formalism, Cartan's calculus of differential forms and the spinor calculus.  In particular, we wish to highlight the compacted spin-coefficient formalism.
It was first developed by Newman and Penrose~\cite{newman_approach_1962} and later streamlined by Geroch, Held and Penrose (GHP), hence the compacted spin-coefficient formalism is also referred to as the GHP formalism~\cite{geroch_space-time_1973}.  Although each formalism makes the same physical predictions, they differ in the ease of their calculations, in particular the spin-coefficient formalism has led to many exact solutions of the Einstein field equations (EFE) which would have been otherwise difficult to find~\cite{stephani_exact_2009}.\\In the present paper, we apply the spin-coefficient formalism to the Bach equations.  Among the most important differences between the Bach equations and the EFE are that they are fourth-order differential equations in the metric as opposed to the second-order EFE, and they are conformally invariant, a consequence of this invariance is that the conformal scale factor is left undetermined by them.  The theory is, however, different to Weyl's~\cite{weyl_gravitation_1918} conformal theory in that the spacetime geometry remains Riemannian and therefore the covariant derivative of the metric is zero (metric-compatibility).\\An attractive feature of Bach's theory from a physical standpoint is that every spacetime locally conformal to an Einstein space (vanishing of traceless Ricci tensor) is a solution of the Bach equations.  Therefore, the physically relevant Schwarzschild, Kerr, gravitational wave and most Friedmann-Lemaitre-Robertson-Walker cosmological spacetimes are also solutions of the Bach equations.  Indeed, Einstein spaces are solutions to the most general action formed from quadratic invariants of the curvature tensors~\cite{lanczos_electricity_1957}.  For further details on the advantages of including higher-order curvature terms in the action, e.g. their stabilizing effects on the divergence structure of gravity, see~\cite{t_hooft_one_1974,stelle_renormalization_1977,stelle_classical_1978}.  Having given some arguments that lend support to the Bach equations as a physical theory, we note that the emphasis of the present paper is solely on the efficiency of the spin-coefficient formalism in solving the Bach equations.\\The main new achievement of this paper is the translation of the Bach equations into compacted spin-coefficient form.  We suggest that this formulation may be used as an efficient alternative to tensor methods for solving the Bach equations.  Arguments in support of this suggestion are the following.  Firstly, the spin-coefficient formalism deals entirely with scalar quantities, which are easily manipulated and may take the form of explicit functions.  Furthermore, because the spin-coefficients are complex they can be represented by 12 quantities, instead of the 24 required in a orthonormal tetrad formalism, or the 40 Christoffel symbols used in coordinate based approaches, therefore fewer terms arise in the calculation.  The previous points, however, are true for any set of tensor or spinor field equations.  The property of the Bach equations that should make the GHP formalism particularly well suited to solving them is their conformal invariance.  Indeed, it has been shown that the GHP formalism incorporates conformal transformations in a straightforward way~\cite{penrose_spinors_1984}.  Another advantage of the GHP formalism is that it allows one to solve directly for the curvature components, thus the Bach equations, considered as fourth-order equations in the metric components, become second-order in the curvature components, which may be easier to solve.  In support of the idea that the GHP formalism may be particularly well suited to solving the Bach equations, we give two straightforward applications of the formalism by solving the Bach equations for a plane-fronted wave spacetime~\cite{ehlers_exact_1962,fiedler_exact_1980,madsen_plane_1990}, and a static spherically symmetric spacetime~\cite{fiedler_exact_1980,mannheim_exact_1989}, where in both examples the general solution for the curvature spinors are obtained in explicit form.\\The outline of the present paper is as follows, in section \ref{bachequations} we introduce the Bach equations in both tensor and spinor form.  In section \ref{spin-coefficientformalism} we introduce the spin-coefficient formalism.  Section \ref{bachequationstranslation} provides the main result of the paper, which is the translation of the Bach equations into the GHP formalism.  In section \ref{applications} we apply the formalism to two simple examples, then in section \ref{conclusion} we summarise our results.  In the \cref{notation,primeasterisk,newmanPenrose,bianchiidentity,cartancalculus} our notation is defined and some useful formulae are provided, for example the prime and asterisk operations, the Newman-Penrose equations, the spinor form of the Bianchi identity and the spin-coefficients in terms of the curls of the null tetrad.

%
\section{The Bach equations}
\label{bachequations}
The tensor form of the Bach equations can be derived from the following action
\begin{equation}
\label{weylsquaredaction}
S = \int C_{abcd}C^{abcd}\sqrt{g}\dd^4x,
\end{equation} 
where $C_{abcd}$ is the Weyl tensor, which is the trace-free part of the Riemann tensor.  They are defined by the vanishing of the Bach tensor, given as follows~\cite{kozameh_conformal_1985}
\begin{equation}
	\label{bachtensor}
		B_{ab} = ({\nabla^c}{\nabla^d} - \tfrac{1}{2}R^{cd})C_{acbd},
\end{equation}
where~$\nabla_a$ is the covariant derivative and~$R_{cd}$ is the Ricci tensor, see~\cref{notation} for further details.
%
%
%
The Bach tensor satisfies the following relations
\begin{equation}
	\label{bachtensorproperties}
		B_{ab} = B_{ba}, \quad g^{ab}B_{ab} = 0, \quad \nabla^aB_{ab} = 0, \quad B_{ab} = \bar{B}_{ab}.
\end{equation}
such that it is symmetric, traceless, divergence-free and real.  In the last relation we have used an over-bar to denote the complex conjugate of a tensor.  Furthermore, under the conformal rescaling
%
\begin{equation}
	\label{metricconformalresclaing}
		g_{ab} \mapsto \hat{g}_{ab} = \Omega^2g_{ab},
\end{equation}
we have
\begin{equation}
	\label{bachconformalinvariance}
		\hat{B}_{ab} = \Omega^{-2}B_{ab},
\end{equation}
such that it is conformally weighted with conformal weight -2~\cite{szekeres_conformal_1968}.  Due to \eqref{bachconformalinvariance}, the vanishing of the Bach tensor is a conformally invariant equation, hence the Bach equations are also conformally invariant.
The Bach spinor is defined as follows\footnote{The ordering of unprimed in relation to primed indices is inconsequential, therefore, it is not necessary to stagger up and down indices of different types.}~\cite{kozameh_conformal_1985}
\begin{equation}
	\label{bachequationsspinor}
	B_{ABA'B'} = 2(\nabla^{C}_{A'}\nabla^{D}_{B'} + \Phi^{CD}_{A'B'})\Psi_{ABCD},
\end{equation}
%
%
where~$\Psi_{ABCD} = \Psi_{(ABCD)}$ is the (complex) Weyl spinor and~$\Phi_{ABC'D'} = \Phi_{(AB)(C'D')} = \conj{\Phi}_{ABC'D'}$ is the (real, traceless) Ricci spinor.  We have also denoted the complex conjugation of a spinor with an over-bar, see \eqref{complexconjugation}.  The Bach spinor has the same symmetries as the Ricci spinor
\begin{equation}
	\label{bachspinorproperties}
	B_{ABA'B'}=B_{(AB)(A'B')}, \quad \bar{B}_{ABC'D'}=B_{ABC'D'}.
\end{equation}
\section{Spin-coefficient formalism}
\label{spin-coefficientformalism}
In this section, in order to translate \eqref{bachequationsspinor} into the spin-coefficient formalism, we introduce a few useful relations.  When using a non-coordinate basis in the tensor calculus, one introduces a tetrad, which is a set of four linearly independent real vector-fields, see \eqref{nulltetraddef}.  The analogous object for the spinor calculus is a dyad, which is a pair of linearly independent complex vector-fields, see \eqref{dyad}.  In terms of the dyad, twelve complex spin-coefficients are defined in the following way
\begin{equation}
	\label{basistospincoefficients}
	\begin{array}{lll}                   
		\kappa = o^ADo_A, & \gamma' = -\iota^ADo_A, & \tau' = -\iota^AD\iota_A, \\        
		\rho = o^A\delta'o_A,  & \beta' = -\iota^A\delta'o_A, & \sigma' = -\iota^A\delta'\iota_A, \\
		\sigma = o^A\delta o_A, &  \beta = \iota^A\delta o_A, & \rho' = -\iota^A\delta\iota_A, \\
		\tau = o^AD'o_A,  & \gamma = \iota^AD'o_A,  & \kappa' = -\iota^AD'\iota_A, \\
	\end{array}                          
\end{equation}
where the \emph{intrinsic derivatives} are the components of the vector covariant derivative in the spinor basis, see \eqref{spinframe}
%
%
\begin{equation}
	\begin{array}{c}
	\label{intrinsicderivative}
		D = o^Ao^{A'}\nabla_{AA'},\\
		\delta = o^A\iota^{A'}\nabla_{AA'},\\
		\delta' = \iota^Ao^{A'}\nabla_{AA'},\\
		D' = \iota^A\iota^{A'}\nabla_{AA'}.
	\end{array}
\end{equation}
The complex conjugate relations of \eqref{basistospincoefficients} and \eqref{intrinsicderivative} follow by replacing the dyad by its complex conjugate and the fact that the covariant derivative is real.  The significance of the primed symbols is that under a discrete transformation called the prime operation, see \eqref{primetransformation}, primed symbols become unprimed and vice versa, thus an economy of notation is achieved.
Let us also define the following standard shorthand, $\xi_{r,t}$, for the components of a symmetric spinor $\xi_{(A\ldots D\ldots)(G'\ldots K')}$~\cite{stephani_exact_2009,penrose_spinors_1984}
\begin{equation}
\label{spinorcomponents}
\xi_{r,t} = \xi_{A\ldots D\ldots G'\ldots K'}\underbrace{o^A..}_{r'} \underbrace{l^D..}_{r} \underbrace{o^{G'}..}_{t'} \underbrace{l^{K'}..}_{t}.
\end{equation}
A particular example of \eqref{spinorcomponents} for the Weyl, Ricci and Bach spinors would be
\begin{equation}
	\label{spinordefinitions}
		\begin{array}{ccc}               
			\Psi_2  = \Psi_{ABCD}\,o^Ao^B\iota^C\iota^D, & \Phi_{11}  = \Phi_{ABA'B'}\,o^A\iota^Bo^{A'}\iota^{B'}, &  B_{21}  = B_{ABA'B'}\,\iota^A\iota^Bo^{A'}\iota^{B'}.
		\end{array}
\end{equation}
Moreover, the Ricci and Bach spinors are real, therefore, we have the following conjugate transpose relations between the components
\begin{equation}
	\label{realriccibach}
		\Phi_{rt}  = \conj{\Phi}_{tr}, \quad B_{rt}  = \conj{B}_{tr}.
\end{equation}
In the spin-coefficient formalism the components of the Weyl and Ricci spinors are related to the intrinsic derivatives of the spin-coefficients via the so called Newman-Penrose equations~\cite{penrose_spinors_1984}.  These relations are given in \cref{newmanPenrose} in compacted form according to the GHP formalism, which we turn to now.
%
\subsection{GHP formalism}
The GHP formalism is a calculus based on a pair of null directions.  It was introduced for situations where two null vectors are physically distinguished by the problem under consideration, but a complete basis is not~\cite{geroch_space-time_1973}.  More relevant to our purposes, is that the compacted expressions are, in the majority of cases, considerably simpler than their counterparts in the original scheme.  This alone justifies our use of the GHP formalism, therefore, the equations may be legitimately regarded as shorthand expressions for full spin-coefficient formulae.\\The most general change of spin-frame which leaves two null directions invariant is
\begin{equation}
	\label{twonullinvariantdyadtransformation}
	o^A \mapsto \lambda o^A, \quad \iota^A \mapsto \lambda^{-1} \iota^A,
\end{equation}
where $\lambda$ is an arbitrary (nowhere vanishing) complex scalar field.  From \eqref{twonullinvariantdyadtransformation} it follows that
\begin{equation}
\label{twonullinvariantdyadtransformationlowerindex}
o_A \mapsto \lambda o_A, \quad \iota_A \mapsto \lambda^{-1} \iota_A,
\end{equation} and that
\begin{equation}
\label{twonullinvariantdyadtransformationprimedindex}
	o^{A'} \mapsto \conj{\lambda} o^{A'}, \quad \iota^{A'} \mapsto \conj{\lambda}^{-1} \iota^{A'}, o_{A'} \mapsto \conj{\lambda} o_{A'}, \quad \iota_{A'} \mapsto \conj{\lambda}^{-1} \iota_{A'}.
\end{equation}
The formalism deals with scalars~$\eta$ associated with a spin-frame or null tetrad where the scalars transform in the following way
\begin{equation}
\label{spinboosttransformation}
	\eta \mapsto \lambda^p\conj{\lambda}^q\eta,
\end{equation}
whenever the dyad transforms as in \eqref{twonullinvariantdyadtransformation}.  A scalar transforming according to \eqref{spinboosttransformation} is called a weighted scalar of type~$(p,q)$.  We have the following weighted scalars
\begin{equation}
	\label{weightedscalars}
		\begin{array}{ll}                   
		\Psi_r & (4 - 2r, 0),\\
		\Phi_{rt} & (2 - 2r, 2 - 2t),\\        
		\Lambda & (0,0),\\
		\rho & (1,1),\\
		\tau & (1,-1),\\
		\kappa & (3,1),\\
		\sigma & (3,-1),\\
		\end{array}                          
\end{equation}
where~$R = 24\Lambda$ is the Ricci scalar, see \eqref{Ricciscalardefinition}.  The remaining weighted scalars are obtained by priming, conjugating or performing both on \eqref{weightedscalars}, where the prime of a~$(p,q)$ scalar is of type~$(-p,-q)$ and the complex conjugate of a~$(p,q)$ scalar is of type~$(q,p)$.
As an example, let us see how~$\sigma$ transforms,
\begin{align}
	\sigma &= o^A\delta o_{A} = o^Ao^B\iota^{B'}\nabla_{BB'}o_{A}\nonumber\\
	&\rightarrow \lambda o^A \lambda o^B {\conj{\lambda}}^{-1} \iota^{B'} \nabla_{BB'}(\lambda o_{A})\nonumber\\
	&={\lambda^2}{\conj{\lambda}}^{-1}o^Ao_Ao^B\iota^{B'}\nabla_{BB'}\lambda + {\lambda^3}{\conj{\lambda}}^{-1}o^Ao^B\iota^{B'}\nabla_{BB'}o_{A}\nonumber\\\label{sigmatransformation}
	&={\lambda^3}{\conj{\lambda}}^{-1}\sigma,
\end{align}
where the first term of the second to last line is zero due to \eqref{spinframe}.
%
\\The spin coefficients of the middle column of \eqref{basistospincoefficients} are not weighted, nor are the intrinsic derivatives \eqref{intrinsicderivative}.  They are therefore combined to give the following weighted derivative operators when acting on a~$(p,q)$ scalar~\cite{geroch_space-time_1973},\footnote{The symbol $\tho$ is pronounced `thorn' and $\edh$ is pronounced `eth'.}
%
\begin{equation}
	\label{edththorn}
	\begin{array}{ll}                   
	\tho = D + p\gamma' + q\conj{\gamma}' & (1,1),\\        
	\tho' = D' - p\gamma - q\conj{\gamma} & (-1,-1),\\
	\edh = \delta - p\beta + q\conj{\beta}' & (1,-1),\\
	\edh' = \delta' + p\beta'  - q\conj{\beta} & (-1,1).
\end{array}                          
\end{equation}
The differential operators~\eqref{edththorn} are of weight~$(p,q)$ in the sense that acting on a scalar of type~$(u,v)$ produces a scalar of type ~$(u + p,v + q)$.  
In terms of \eqref{edththorn}, the following relations follow from \eqref{basistospincoefficients}
\begin{equation}
\label{compactedbasistospincoefficientsandintrinsicderivatives}
\begin{array}{ll}                   
	\tho o^A = - \kappa l^A, & \tho l^A = - \tau'o^A,\\        
	\edh'o^A = - \rho l^A, & \edh'l^A = - \sigma'o^A,\\        
	\edh o^A = - \sigma l^A, & \edh l^A = - \rho'o^A,\\   
	\tho'o^A = - \tau l^A, & \tho'l^A = - \kappa'o^A.
\end{array}                          
\end{equation}
%
%
Let~$\xi_{A\ldots D\ldots G'\ldots K'}$ be a symmetric spinor with components defined by \eqref{spinorcomponents}.
Making use of \eqref{compactedbasistospincoefficientsandintrinsicderivatives} and their complex conjugate relations, the corresponding components of the intrinsic derivatives of $\xi_{A\ldots D\ldots G'\ldots K'}$ may be derived
\begin{align}
\begin{split}
	\label{intrinsicderivativetocompactedintrinsicderivative}                   
	(o^A \ldots l^D \ldots o^{G'} \ldots l^{K'} \ldots)D\xi_{A\ldots D\ldots G'\ldots K'} =&\tho\xi_{r,t} + r'\kappa\xi_{r + 1,t} + r\tau'\xi_{r - 1,t}\\&+ t'\conj{\kappa}\xi_{r,t + 1} + t\conj{\tau}'\xi_{r,t - 1},\\
	(o^A \ldots l^D \ldots o^{G'} \ldots l^{K'} \ldots)\delta\xi_{A\ldots D\ldots G'\ldots K'} =&\edh\xi_{r,t} + r'\sigma\xi_{r + 1,t} + r\rho'\xi_{r - 1,t}\\&+ t'\conj{\rho}\xi_{r,t + 1} + t\conj{\sigma}'\xi_{r,t - 1},\\
	(o^A \ldots l^D \ldots o^{G'} \ldots l^{K'} \ldots)\delta'\xi_{A\ldots D\ldots G'\ldots K'} =&\edh'\xi_{r,t} + r'\rho\xi_{r + 1,t} + r\sigma'\xi_{r - 1,t}\\&+ t'\conj{\sigma}\xi_{r,t + 1} + t\conj{\rho}'\xi_{r,t - 1},\\
	(o^A \ldots l^D \ldots o^{G'} \ldots l^{K'} \ldots)D'\xi_{A\ldots D\ldots G'\ldots K'} =&\tho'\xi_{r,t} + r'\tau\xi_{r + 1,t} + r\kappa'\xi_{r - 1,t}\\&+ t'\conj{\tau}\xi_{r,t + 1} + t\conj{\kappa}'\xi_{r,t - 1}.
\end{split}
\end{align}
%
%
%
%
Equations \eqref{intrinsicderivativetocompactedintrinsicderivative} are the key equations which provide an efficient way of translating differential spinor equations into the GHP formalism.  All the formulae hitherto are known and can be found in the literature, e.g. see~\cite{penrose_spinors_1984}.
\section{Bach equations in terms of spin-coefficients}
\label{bachequationstranslation}
With the help of \eqref{intrinsicderivativetocompactedintrinsicderivative} the translation of \eqref{bachequationsspinor} into spin-coefficient form is straightforward.  According to \eqref{dyadcomponents} and its complex conjugate relation we first take components of \eqref{bachequationsspinor} as follows
\begin{equation}
	\label{bachequationsspinorcomponents}
		\tfrac{1}{2}B_{\textbf{AB}\textbf{A}'\textbf{B}'} = -{\conj{\varepsilon}_{\textbf{B}'}}^{B'}{\varepsilon_{\textbf{A}}}^{A}{\varepsilon_{\textbf{B}}}^{B}{\varepsilon_{\textbf{C}}}^{C}\nabla^{\textbf{C}}_{\textbf{A}'}\nabla^{D}_{B'}\Psi_{ABCD} - \Phi^{\textbf{CD}}_{\textbf{A}'\textbf{B}'}\Psi_{\textbf{ABCD}},
\end{equation}
where bold font captial indices refer to a spinor basis, see \cref{notation}.  Writing out the summation terms in the second term is straightforward.  For the first term, we write out the summation terms which include the components of the outer covariant derivative $\nabla^{\textbf{C}}_{\textbf{A}'}$ acting on the spinor $\nabla^{D}_{B'}\Psi_{ABCD}$.  We find terms of the form of the left hand side of \eqref{intrinsicderivativetocompactedintrinsicderivative} with $\nabla^{D}_{B'}\Psi_{ABCD}$ playing the role of $\xi_{A\ldots D\ldots G'\ldots K'}$.  Because it is symmetric, we may substitute the right hand side of \eqref{intrinsicderivativetocompactedintrinsicderivative} with the corresponding values of $r, r', t, t'$.  We then repeat this procedure for the inner covariant derivative, where now the spinor $\Psi_{ABCD}$ plays the role of $\xi_{A\ldots D\ldots G'\ldots K'}$. 
By this two-step procedure we obtain an expression solely in terms of the derivative operators and spin-coefficients of the GHP formalism.  The Bach tensor in spin-coefficient form is then given as follows,\footnote{We adhere to the convention that a differential operator acts only on the symbol (or bracketed expression) which immediately follows it-unless this is also a differential operator.}
%
%
%
\begin{subequations}
	\label{bachspinorcomponents4group}
	\begin{align}
	\label{B00bachspinor}
	\tfrac{1}{2}B_{00} = {}&(\tho - 3\rho)[(\edh' - 2\tau')\Psi_1 - (\tho - 3\rho)\Psi_2 + \sigma'\Psi_0 - 2\kappa\Psi_3] \\ \nonumber &+ (\edh' - \tau')[(\tho - 4\rho)\Psi_1 - (\edh' - \tau')\Psi_0 + 3\kappa\Psi_2] \\ \nonumber
	&+ 2\kappa[(\edh' - 3\tau')\Psi_2 - (\tho - 2\rho)\Psi_3 + 2\sigma'\Psi_1 - \kappa\Psi_4] \\ \nonumber
	&+ \conj{\kappa}[(\tho' - 2\rho')\Psi_1 - (\edh - 3\tau)\Psi_2 + \kappa'\Psi_0 - 2\sigma\Psi_3] \\ \nonumber
	&+ \conj{\sigma}[(\edh - 4\tau)\Psi_1 - (\tho' - \rho')\Psi_0 + 3\sigma\Psi_2] - \Phi_{20}\Psi_0 + 2\Phi_{10}\Psi_1 - \Phi_{00}\Psi_2, \\
	\label{B02bachspinor}
	\tfrac{1}{2}B_{02} = {}&(\edh - 3\tau)[(\tho' - 2\rho')\Psi_1 - (\edh - 3\tau)\Psi_2 + \kappa'\Psi_0 - 2\sigma\Psi_3] \\ \nonumber &+ (\tho' - \rho')[(\edh - 4\tau)\Psi_1 - (\tho' - \rho')\Psi_0 + 3\sigma\Psi_2] \\ \nonumber
	&+ 2\sigma[(\tho' - 3\rho')\Psi_2 - (\edh - 2\tau)\Psi_3 + 2\kappa'\Psi_1 - \sigma\Psi_4] \\ \nonumber
	&+ \conj{\sigma}'[(\edh' - 2\tau')\Psi_1 - (\tho - 3\rho)\Psi_2 + \sigma'\Psi_0 - 2\kappa\Psi_3] \\ \nonumber
	&+ \conj{\kappa}'[(\tho - 4\rho)\Psi_1 - (\edh' - \tau')\Psi_0 + 3\kappa\Psi_2] - \Phi_{22}\Psi_0 + 2\Phi_{12}\Psi_1 - \Phi_{02}\Psi_2, \\
	\label{B20bachspinor}
	\tfrac{1}{2}B_{20} = {}&(\edh' - 3\tau')[(\tho - 2\rho)\Psi_3 - (\edh' - 3\tau')\Psi_2 + \kappa\Psi_4 - 2\sigma'\Psi_1] \\ \nonumber &+ (\tho - \rho)[(\edh' - 4\tau')\Psi_3 - (\tho - \rho)\Psi_4 + 3\sigma'\Psi_2] \\ \nonumber
	&+ 2\sigma'[(\tho - 3\rho)\Psi_2 - (\edh' - 2\tau')\Psi_1 + 2\kappa\Psi_3 - \sigma'\Psi_0] \\ \nonumber
	&+ \conj{\sigma}[(\edh - 2\tau)\Psi_3 - (\tho' - 3\rho')\Psi_2 + \sigma\Psi_4 - 2\kappa'\Psi_1] \\ \nonumber
	&+ \conj{\kappa}[(\tho' - 4\rho')\Psi_3 - (\edh - \tau)\Psi_4 + 3\kappa'\Psi_2] - \Phi_{00}\Psi_4 + 2\Phi_{10}\Psi_3 - \Phi_{20}\Psi_2, \\
	\label{B22bachspinor}
	\tfrac{1}{2}B_{22} = {}&(\tho' - 3\rho')[(\edh - 2\tau)\Psi_3 - (\tho' - 3\rho')\Psi_2 + \sigma\Psi_4 - 2\kappa'\Psi_1] \\ \nonumber &+ (\edh - \tau)[(\tho' - 4\rho')\Psi_3 - (\edh - \tau)\Psi_4 + 3\kappa'\Psi_2] \\ \nonumber
	&+ 2\kappa'[(\edh - 3\tau)\Psi_2 - (\tho' - 2\rho')\Psi_1 + 2\sigma\Psi_3 - \kappa'\Psi_0] \\ \nonumber
	&+ \conj{\kappa}'[(\tho - 2\rho)\Psi_3 - (\edh' - 3\tau')\Psi_2 + \kappa\Psi_4 - 2\sigma'\Psi_1] \\ \nonumber
	&+ \conj{\sigma}'[(\edh' - 4\tau')\Psi_3 - (\tho - \rho)\Psi_4 + 3\sigma'\Psi_2] - \Phi_{02}\Psi_4 + 2\Phi_{12}\Psi_3 - \Phi_{22}\Psi_2,
	\end{align}
\end{subequations}
\begin{subequations}
	\label{bachspinorcomponents2agroup}
	\begin{align}
	\label{B01bachspinor}
	\tfrac{1}{2}B_{01} = {}&(\tho - 3\rho)[(\tho' - 2\rho')\Psi_1 - (\edh - 3\tau)\Psi_2 + \kappa'\Psi_0 - 2\sigma\Psi_3] \\ \nonumber &+ (\edh' - \tau')[(\edh - 4\tau)\Psi_1 - (\tho' - \rho')\Psi_0 + 3\sigma\Psi_2] \\ \nonumber
	&+ 2\kappa[(\tho' - 3\rho')\Psi_2 - (\edh - 2\tau)\Psi_3 + 2\kappa'\Psi_1 - \sigma\Psi_4] \\ \nonumber
	&+ \conj{\tau}'[(\edh' - 2\tau')\Psi_1 - (\tho - 3\rho)\Psi_2 + \sigma'\Psi_0 - 2\kappa\Psi_3] \\ \nonumber
	&+ \conj{\rho}'[(\tho - 4\rho)\Psi_1 - (\edh' - \tau')\Psi_0 + 3\kappa\Psi_2] - \Phi_{21}\Psi_0 + 2\Phi_{11}\Psi_1 - \Phi_{01}\Psi_2, \\
	\label{B21bachspinor}
	\tfrac{1}{2}B_{21} = {}&(\tho' - 3\rho')[(\tho - 2\rho)\Psi_3 - (\edh' - 3\tau')\Psi_2 + \kappa\Psi_4 - 2\sigma'\Psi_1] \\ \nonumber &+ (\edh - \tau)[(\edh' - 4\tau')\Psi_3 - (\tho - \rho)\Psi_4 + 3\sigma'\Psi_2] \\ \nonumber
	&+ 2\kappa'[(\tho - 3\rho)\Psi_2 - (\edh' - 2\tau')\Psi_1 + 2\kappa\Psi_3 - \sigma'\Psi_0] \\ \nonumber
	&+ \conj{\tau}[(\edh - 2\tau)\Psi_3 - (\tho' - 3\rho')\Psi_2 + \sigma\Psi_4 - 2\kappa'\Psi_1] \\ \nonumber
	&+ \conj{\rho}[(\tho' - 4\rho')\Psi_3 - (\edh - \tau)\Psi_4 + 3\kappa'\Psi_2] - \Phi_{01}\Psi_4 + 2\Phi_{11}\Psi_3 - \Phi_{21}\Psi_2,
		\end{align}
\end{subequations}
\begin{subequations}
	\label{bachspinorcomponents2bgroup}
	\begin{align}
	\label{B10bachspinor}
	\tfrac{1}{2}B_{10} = {}&(\tho - 2\rho)[(\edh' - 3\tau')\Psi_2 - (\tho - 2\rho)\Psi_3 + 2\sigma'\Psi_1 - \kappa\Psi_4] \\ \nonumber &+ (\edh' - 2\tau')[(\tho - 3\rho)\Psi_2 - (\edh' - 2\tau')\Psi_1  - \sigma'\Psi_0 + 2\kappa\Psi_3] \\ \nonumber
	&+ \kappa[(\edh' - 4\tau')\Psi_3 - (\tho - \rho)\Psi_4 + 3\sigma'\Psi_2] \\ \nonumber
	&+ \conj{\kappa}[(\tho' - 3\rho')\Psi_2 - (\edh - 2\tau)\Psi_3 + 2\kappa'\Psi_1 - \sigma\Psi_4] \\ \nonumber
	&+ \conj{\sigma}[(\edh - 3\tau)\Psi_2 - (\tho' - 2\rho')\Psi_1 - \kappa'\Psi_0 + 2\sigma\Psi_3] - \Phi_{20}\Psi_1 + 2\Phi_{10}\Psi_2 - \Phi_{00}\Psi_3, \\
	\label{B12bachspinor}
	\tfrac{1}{2}B_{12} = {}&(\tho' - 2\rho')[(\edh - 3\tau)\Psi_2 - (\tho' - 2\rho')\Psi_1 + 2\sigma\Psi_3 - \kappa'\Psi_0] \\ \nonumber &+ (\edh - 2\tau)[(\tho' - 3\rho')\Psi_2 - (\edh - 2\tau)\Psi_3  - \sigma\Psi_4 + 2\kappa'\Psi_1] \\ \nonumber
	&+ \kappa'[(\edh - 4\tau)\Psi_1 - (\tho' - \rho')\Psi_0 + 3\sigma\Psi_2] \\ \nonumber
	&+ \conj{\kappa}'[(\tho - 3\rho)\Psi_2 - (\edh' - 2\tau')\Psi_1 + 2\kappa\Psi_3 - \sigma'\Psi_0] \\ \nonumber
	&+ \conj{\sigma}'[(\edh' - 3\tau')\Psi_2 - (\tho - 2\rho)\Psi_3 + 2\sigma'\Psi_1 - \kappa\Psi_4] - \Phi_{02}\Psi_3 + 2\Phi_{12}\Psi_2 - \Phi_{22}\Psi_1,
	\end{align}
\end{subequations}
\begin{align}
\label{B11bachspinor}
\tfrac{1}{2}B_{11} = {}&(\tho - 2\rho)[(\tho' - 3\rho')\Psi_2 - (\edh - 2\tau)\Psi_3 + 2\kappa'\Psi_1 - \sigma\Psi_4] \\ \nonumber &+ (\edh' - 2\tau')[(\edh - 3\tau)\Psi_2 - (\tho' - 2\rho')\Psi_1  - \kappa'\Psi_0 + 2\sigma\Psi_3] \\ \nonumber
&+ \kappa[(\tho' - 4\rho')\Psi_3 - (\edh - \tau)\Psi_4 + 3\kappa'\Psi_2] \\ \nonumber
&+ \sigma'[(\edh - 4\tau)\Psi_1 - (\tho' - \rho')\Psi_0 + 3\sigma\Psi_2] \\ \nonumber
&+ \conj{\rho}'[(\tho - 3\rho)\Psi_2 - (\edh' - 2\tau')\Psi_1 + 2\kappa\Psi_3 - \sigma'\Psi_0] - \Phi_{21}\Psi_1 + 2\Phi_{11}\Psi_2 - \Phi_{01}\Psi_3.
\end{align}
%
%
Setting equations \eqref{bachspinorcomponents4group}, \eqref{bachspinorcomponents2agroup}, \eqref{bachspinorcomponents2bgroup} and \eqref{B11bachspinor} to zero gives the Bach equations in spin-coefficient form.  In addition to the prime operation there exists another discrete transformation effected by the asterisk operator (*).  The Bach equations in the GHP formalism are interchanged by the prime and asterisk operations respectively in the following way,
%
\begin{equation}
\label{bachprime}
	\begin{array}{lll}                   
	B_{rs} \mapsto B_{tu} & 0 \leftrightarrow 2 & 1 \leftrightarrow 1,
\end{array}                          
\end{equation}
\begin{equation}
\label{asteriskoperation}                     
	\begin{array}{l}                         
	B_{01}  \leftrightarrow -B_{01}, \\
	B_{21}  \leftrightarrow -B_{21}, \\
	B_{10}  \leftrightarrow B_{12}, \\
	B_{00}  \leftrightarrow B_{02}, \\
	B_{22}  \leftrightarrow B_{20}, \\
	B_{11}  \leftrightarrow -B_{11}.
\end{array}
\end{equation}
Therefore, the Bach equations split into the four groups \eqref{bachspinorcomponents4group}, \eqref{bachspinorcomponents2agroup}, \eqref{bachspinorcomponents2bgroup} and \eqref{B11bachspinor} which under the combined action of \eqref{bachprime} and \eqref{asteriskoperation} transform only amongst themselves, see \cref{primeasterisk} for the general definition of these transformations.  
\section{Applications of the Bach equations in spin-coefficient form}
\label{applications}
\subsection{PP-wave spacetime}
As a straightforward example we solve the Bach equations for a metric corresponding to a pp-wave spacetime,\footnote{A plane-fronted gravitational wave with parallel rays spacetime, abbreviated as a \emph{pp-wave} spacetime, was defined in~\cite{ehlers_exact_1962} to be any Lorentzian manifold which admits a covariantly constant null vector field.  Furthermore, the authors showed that a metric for such a spacetime can always be written in the form~\eqref{planewavemetric}, which was investigated earlier, see~\cite{brinkmann_einstein_1925}.} 
which may be represented in a coordinate chart~$(u,v,z,\conj{z})$ by the following fundamental form  
%
\begin{equation}
	\label{planewavemetric}
	g_{\textbf{ab}}\dd x^\textbf{a} \dd x^\textbf{b} = 2H\dd u^2 + (\dd u\,\dd v + \dd v\,\dd u) - (\dd z\,\dd \conj{z} + \dd \conj{z}\,\dd z),
\end{equation}
where~$H=H(u,z,\conj{z})$.
%
%
%
By substituting the components of the metric from \eqref{planewavemetric} into the component form of \eqref{normalisednulltetrad} and solving for ${e_\textbf{m}}^\textbf{a}$, we find the following 
%
%
%
%
\begin{equation}
	\label{planewavenulltetradmatrix}
		\begin{pmatrix} 
		l^\textbf{a} \\
		n^\textbf{a} \\
		m^\textbf{a} \\
		\conj{m}^\textbf{a}
		\end{pmatrix} =
			\begin{pmatrix}
				0 & 1 & 0 & 0 \\
				1 & -H & 0 & 0 \\
				0 & 0 & 1 & 0 \\
				0 & 0 & 0 & 1
			\end{pmatrix},
\end{equation}
where a normalisation choice was made.  Lowering the coordinate basis index gives
%
%
%
%
\begin{equation}
\label{nulltetrad}
	\begin{pmatrix} 
	l_\textbf{a} \\
	n_\textbf{a} \\
	m_\textbf{a} \\
	\conj{m}_\textbf{a}
	\end{pmatrix} =
	\begin{pmatrix}
	1 & 0 & 0 & 0 \\
	H & 1 & 0 & 0 \\
	0 & 0 & 0 & -1 \\
	0 & 0 & -1 & 0
	\end{pmatrix}.
\end{equation}
%
%
Substituting \eqref{planewavenulltetradmatrix} and \eqref{nulltetrad} into \eqref{solutionlnmbasisvectors} we find the following non-zero spin-coefficient for a pp-wave spacetime 
%
%
%
%
\begin{equation}
	\label{planewavenonzerospincoefficients}
\kappa'	= -H_{,\conj{z}},
\end{equation}
%
%
where the comma stands for the partial derivative, all other spin-coefficients are zero.
Substituting \eqref{planewavenonzerospincoefficients} into \eqref{compactednpequations}, we find the following non-zero curvature spinor components
\begin{align}
%
%
%
%
	\label{planewaveweylsolution}
		\Psi_{4} & = -\delta'\kappa' = H_{,\conj{z}\conj{z}},\\	
	\label{planewavericcisolution}
		\Phi_{22} & = -\delta\kappa' = H_{,\conj{z}z},
\end{align}
all other curvature spinor components are zero.
From \eqref{planewavenonzerospincoefficients}, \eqref{planewaveweylsolution} and \eqref{planewavericcisolution} we find that all the Bach equations are trivially satisfied except for \eqref{B22bachspinor} which gives
\begin{equation}
	\label{nontrivialbachequationsspinor}
	\edh^2\Psi_{4} = \delta^2\Psi_{4} = {\Psi_{4}}_{,zz} = 0.
\end{equation}
%
%
%
%
%
The general solution to \eqref{nontrivialbachequationsspinor} is
\begin{equation}
	\label{weylspinorsolutioncomplexcoordinates}
		\Psi_{4}(u, z, \conj{z}) = z\psi_{1}(u, \conj{z}) + \psi_{2}(u,\conj{z}),
\end{equation}
where~$\psi_{1}(u, \conj{z})$ and~$\psi_{2}(u,\conj{z})$ are arbitrary complex functions.  From \eqref{planewaveweylsolution} and \eqref{planewavericcisolution} we find
\begin{equation}
	\label{planewavebianchiidentity}
	{\Psi_{4}}_{,z} = {\Phi_{22}}_{,\conj{z}},
\end{equation}
which is the only non-trivial Bianchi identity remaining for a pp-wave spacetime, cf. \eqref{compactedbianchiidentityspinor}.  Substituting \eqref{weylspinorsolutioncomplexcoordinates} into \eqref{planewavebianchiidentity} we find the general solution for the Ricci spinor
\begin{equation}
	\label{riccispinorsolutioncomplexcoordinates}
	\Phi_{22}(u, z, \conj{z}) = \phi(u, z) + \conj{\phi}(u,\conj{z}),
\end{equation}
where
\begin{equation}
	\label{planewaveweylriccirelation}
	\conj{\phi}_{,\conj{z}} = \psi_{1}.
\end{equation}
Equations \eqref{weylspinorsolutioncomplexcoordinates} and \eqref{riccispinorsolutioncomplexcoordinates} are the general solutions to a special case of the examples considered in~\cite{madsen_plane_1990,fiedler_exact_1980}.  The function $H$ can be obtained implicitly by integrating twice either \eqref{weylspinorsolutioncomplexcoordinates} or \eqref{riccispinorsolutioncomplexcoordinates} according to \eqref{planewaveweylsolution} or \eqref{planewavericcisolution} respectively.
\\For pp-waves, the physical interpretation of the curvature components $\Psi_{4}$ and $\Phi_{22}$ is the following.  The modulus and argument of~$\Psi_{4}$ correspond respectively to the amplitude and polarization of the gravitational plane wave, and in Einstein-Maxwell theory the square root of~$\Phi_{22}$ corresponds to the electromagnetic part of the wave.  The polarization of the electromagnetic part, whilst still an arbitrary function of~$u$, does not contribute to the curvature~\cite{penrose_remarkable_1965}.  A particularly simple type of pp-wave is a \emph{plane wave}.  In this case the function $H$ depends on $z$ and $\conj{z}$ in the following way~\cite{penrose_remarkable_1965,stephani_exact_2009}
\begin{equation}
\label{planewaveH}
H(u,z,\conj{z})= A(u)z^2 + \conj{A}(u)\conj{z}^2 + B(u)z\conj{z},
\end{equation}
where $A$ is complex and $B$ is real.
%
%
%
Substituting \eqref{planewaveH} into  \eqref{planewaveweylsolution} and \eqref{planewavericcisolution} we find
\begin{align}
\label{planewaveweylsolutionquadratic}
\Psi_{4} & = A,\\
\label{planewavericcisolutionquadratic}
\Phi_{22} & = B.
\end{align}
In this case \eqref{nontrivialbachequationsspinor} is trivially satisfied since~$\Psi_4$ depends only on~$u$.  A different particular case is an Einstein space, defined by the equation $\Phi_{ab} = 0$, corresponding to a pure gravitational wave.  From \eqref{riccispinorsolutioncomplexcoordinates} and \eqref{planewaveweylriccirelation} we have $\phi = 0$ and $\psi_{1}=0$.  Therefore, from \eqref{weylspinorsolutioncomplexcoordinates} the gravitational wave has amplitude and polarization equal to the modulus and argument of $\psi_{2}$ respectively.
\subsection{Static spherically symmetric spacetime}
%
For our second example we consider a static spherically symmetric spacetime, i.e. the conditions under which the Schwarzschild solution is the unique solution of the EFE.  The solution to the Bach equations under these conditions was found in~\cite{mannheim_exact_1989}, where the authors show that a static spherically symmetric spacetime may be represented in a coordinate chart~$(t,r,\theta,\phi)$ by the following fundamental form
%
\begin{equation}
\label{sphericallysymmetricmetric}
g_{\textbf{ab}}\dd x^{\textbf{a}} \dd x^{\textbf{b}}  = \frac{p^2}{r^2}\left(A^2\dd t^2 - A^{-2}\dd r^2 - r^2 \dd \theta^2 - r^2 \sin^2\theta\,\dd \phi^2 \right),
\end{equation}
where~$p=p(r)$ and~$A=A(r)$.  Since the Bach equations \eqref{bachequationsspinor} are conformally invariant we may make the following conformal rescaling
\begin{equation}
\label{metricconformalresclaingexample}
	g_{\textbf{ab}} \mapsto \hat{g}_{\textbf{ab}} = \frac{r^2}{p^2}g_{\textbf{ab}},
\end{equation}
such that our metric is transformed to
\begin{equation}
\label{rescaledsphericallysymmetricmetric}
	\hat{g}_{\textbf{ab}}\dd x^{\textbf{a}} \dd x^{\textbf{b}} = A^2\dd t^2 - A^{-2}\dd r^2 - r^2 \dd \theta^2 - r^2 \sin^2\theta\,\dd \phi^2.
\end{equation}
All quantities (e.g. spin-coefficients and spinor components) will now refer to the conformally rescaled metric \eqref{rescaledsphericallysymmetricmetric}, however, we omit the hats on these quantities.
%
%
%
%
%
Substituting the components of \eqref{rescaledsphericallysymmetricmetric} into the component form of \eqref{normalisednulltetrad} we find
\begin{equation}
\label{sphericallysymmetrictetradmatrix}
	\begin{pmatrix} 
	l^\textbf{a} \\
	n^\textbf{a} \\
	m^\textbf{a} \\
	\conj{m}^\textbf{a}
	\end{pmatrix} = \frac{1}{\sqrt{2}}
	\begin{pmatrix}
	A^{-1} & -A & 0 & 0 \\
	A^{-1} & A & 0 & 0 \\
	0 & 0 & -r^{-1} & -ir^{-1}\mathrm{cosec}\theta \\
	0 & 0 & -r^{-1} & ir^{-1}\mathrm{cosec}\theta
\end{pmatrix}.
\end{equation}
Lowering the coordinate basis index gives
\begin{equation}
\label{sphericallysymmetrictetrad}
	\begin{pmatrix} 
	l_\textbf{a} \\
	n_\textbf{a} \\
	m_\textbf{a} \\
	\conj{m}_\textbf{a}
	\end{pmatrix} = \frac{1}{\sqrt{2}}
	\begin{pmatrix}
	A & A^{-1} & 0 & 0 \\
	A & -A^{-1} & 0 & 0 \\
	0 & 0 & r & ir\sin\theta \\
	0 & 0 & r & -ir\sin\theta
	\end{pmatrix}.
\end{equation}
%
Substituting \eqref{sphericallysymmetrictetradmatrix} and \eqref{sphericallysymmetrictetrad} into \eqref{solutionlnmbasisvectors} we calculate the following spin-coefficients, cf.~\cite{davis_simple_1976-1},
\begin{subequations}
	\label{sphericalspincoefficients}
	\begin{align}
		\rho &= -\rho' = \frac{A}{\sqrt{2}r},\label{sphericalspincoefficients1}\\
		\gamma &= -\gamma' = -\frac{\dot{A}}{2\sqrt{2}},\label{sphericalspincoefficients2}\\
		\beta &= -\beta' = -\frac{\cot\theta}{2\sqrt{2}r},\label{sphericalspincoefficients3}
	\end{align}
\end{subequations}
where we have denoted differentiation with respect to the radial coordinate~$r$, not the time, with an over-dot because the prime is already in use.  All other spin-coefficients are zero.  Substituting \eqref{sphericalspincoefficients} into \eqref{compactednpequations} and \eqref{compactednpcommutators} we find that the Newman-Penrose equations and commutator expressions reduce to the following
\begin{subequations}
	\label{sphericalnpeqns}
		\begin{align}
		-2r\rho\dot{\gamma} &= -4\gamma^2 + \Psi_{2} + \Phi_{11} - \Lambda,\label{sphericalnpeqns1}\\
		4\gamma\rho &= \Psi_{2} + 2\Lambda,\label{sphericalnpeqns2}\\
		\frac{1}{2r^2} &= \rho^2 - \Psi_{2} + \Phi_{11} + \Lambda,\label{sphericalnpeqns3}
		\end{align}
\end{subequations}
where $\conj{\Psi}_{2} = \Psi_{2}$ and all other curvature spinor components are zero.  From \eqref{sphericalspincoefficients} and \eqref{sphericalnpeqns} we find that all the Bach equations are trivially satisfied except for \eqref{B00bachspinor} and \eqref{B11bachspinor}.  The former is given by
%
%
%
\begin{equation}
\label{b00conformalequationsimple}
	(\tho - 3\rho)\left[(\tho - 3\rho)\Psi_{2}\right] = 0.
\end{equation}
%
From \eqref{edththorn}, \eqref{sphericallysymmetrictetrad}, \eqref{sphericalspincoefficients} and \eqref{compactednpequations}, we find that \eqref{b00conformalequationsimple} reduces to
%
\begin{equation}
\label{b00conformalequationsimpleexplicit}
	\rho^2\left[r^2\ddot{\Psi}_{2} + 6(r\dot{\Psi}_{2} + \Psi_{2})\right] = 0.
\end{equation}
If $\rho = 0$ then from \eqref{sphericalspincoefficients1} and \eqref{sphericallysymmetricmetric} we see that this solution trivially corresponds to a singular metric tensor.  Therefore, we must have
\begin{equation}
\label{b00conformalequationsimpleexplicitwithoutrho}
r^2\ddot{\Psi}_{2} + 6(r\dot{\Psi}_{2} + \Psi_{2}) = 0,
\end{equation}
from which we obtain the general solution
\begin{equation}
	\label{psi2generalsol}
	\Psi_{2}(r) = \frac{c_1}{r^2} + \frac{c_2}{r^3},
\end{equation}
where~$c_1$ and~$c_2$ are constants of integration.  Equation \eqref{B11bachspinor} is given by the following
%
%
%
\begin{equation}
\label{b11conformalequationsimple}
(\tho - 2\rho)((\tho' - 3\rho')\Psi_{2}) + \rho'(\tho - 3\rho)\Psi_{2} +2\Phi_{11}\Psi_{2} = 0.
\end{equation}
Using \eqref{b00conformalequationsimple} we may further simply \eqref{b11conformalequationsimple},\footnote{Note that $\tho\tho'\Psi_{2} = -\tho^2\Psi_{2} - 4\gamma\tho\Psi_{2}$ and $\tho\rho'=-\tho\rho - 4\gamma\rho$.
}
\begin{equation}
\label{b11conformalequationsimpleexplicit}
(2\gamma\rho + \rho^2)\left(r\dot{\Psi}_{2} + 3\Psi_{2}\right) + \Phi_{11}\Psi_{2} = 0.
\end{equation}
Equations \eqref{b11conformalequationsimpleexplicit} and \eqref{sphericalnpeqns} are the remaining equations that we need to solve.  To achieve this we make use of the Bianchi identities \eqref{compactedbianchiidentityspinor3} and \eqref{compactedbianchiidentityspinor11} which, in  the case of a static spherically symmetric spacetime, are given by
\begin{align}
	(\tho - 3\rho)\Psi_{2} - 2\rho\Phi_{11} + 2\tho\Lambda &= 0,	\label{sphericalbianchiidentitity1}\\
	(\tho - 4\rho)\Phi_{11} + 3\tho\Lambda &= 0.\label{sphericalbianchiidentitity2}
\end{align}
Assuming $\rho\neq0$, \eqref{sphericalbianchiidentitity1} and \eqref{sphericalbianchiidentitity2} reduce to the following
\begin{align}
	r\dot{\Psi}_{2} + 3\Psi_{2} + 2\Phi_{11} + 2r\dot{\Lambda} &= 0,\label{sphericalbianchiidentityrelationsimple1}\\
	r\dot{\Phi}_{11} + 4\Phi_{11} + 3r\dot{\Lambda} &= 0.\label{sphericalbianchiidentityrelationsimple2}
\end{align}
%
%
Substituting \eqref{psi2generalsol} into \eqref{sphericalbianchiidentityrelationsimple1}, we solve \eqref{sphericalbianchiidentityrelationsimple1} and \eqref{sphericalbianchiidentityrelationsimple2} yielding
%
%
\begin{align}
\Phi_{11} &= -\frac{3c_{1}}{2r^2} + \frac{c_{3}}{r},\label{sphericalphi11}\\
\Lambda &= -\frac{c_{1}}{2r^2} + \frac{c_{3}}{r} - c_4,\label{sphericallambda}
\end{align}
where $c_3$ and  $c_4$ are constants of integration.   
%
Substituting \eqref{psi2generalsol}, \eqref{sphericalphi11} and \eqref{sphericallambda} into \eqref{sphericalnpeqns3} we obtain~$\rho^2$, hence from~\eqref{sphericalspincoefficients1} we find~$A^2$ to be
\begin{equation}
\label{sphericalmetricsolution}
A^2 = 1 + 6c_{1} + \frac{2c_{2}}{r} - 4c_{3}r + 2c_4 r^2.
\end{equation}
We find~$\gamma$ from \eqref{sphericalspincoefficients2} and substitute it along with~\eqref{psi2generalsol}, \eqref{sphericalphi11} and \eqref{sphericalmetricsolution} into \eqref{b11conformalequationsimpleexplicit} to obtain the following relation between our constants of integration,
\begin{equation}
\label{constantrelation}
3{c_{1}}^2 + c_{1} + 2c_{2}c_{3} = 0.
\end{equation}
The third term of \eqref{sphericalmetricsolution} is present in the \emph{Schwarzschild} solution where $c_2$ would be equal to (minus) the Newtonian mass~\cite{chandrasekhar_mathematical_1998}.  The fifth term is the cosmological constant term present in the \emph{De Sitter-Schwarzschild} solution.  The second and fourth term, however, are peculiar to the solution of the Bach equations, (cf.~\cite{mannheim_exact_1989} where $c_3$ is proportional to their constant $\gamma$).  
\section{Conclusion}
\label{conclusion}
Since there exist very few exact solutions to the Bach equations in the literature, we hope that our translation of the Bach equations into spin-coefficient form will be useful for finding new exact solutions that would be otherwise difficult to obtain.  This hope is sustained by the fact that the spin-coefficient formalism has already proven to be a powerful method for finding exact solutions to the Einstein field equations~\cite{stephani_exact_2009}.\\In order to show the efficiency of the formalism applied to the Bach equations, we chose two straightforward examples, the plane-fronted wave spacetime and a static spherically symmetric spacetime.  In comparison to coordinate based tensor methods, the calculations involved are shorter.  For example, in a standard approach, one would start with the metric, from which the 40 Christoffel symbols are derived, from these the Riemann tensor follows and its trace and trace-free parts give the Ricci and Weyl tensors.  These are then substituted into the Bach equations yielding fourth-order partial differential equations in the metric components.  On the other hand, starting from the Bach equations in spin-coefficient form, the steps are far fewer, as shown in our examples.  Furthermore, the equations are never more than second-order, since one solves directly for the curvature components as opposed to the metric components.  In more complicated problems we expect the improved efficiency to increase.
\appendix
\renewcommand{\thesection}{\Alph{section}}
\numberwithin{equation}{section}
\section{Tensor and Spinor Calculus notation}
\label{notation}
We adopt the abstract index approach, whereby non-bold font indices are abstract markers and do not take on numerical values.  Whereas bold font indices do take on numerical values and therefore refer to a basis, e.g. a null tetrad or a spinor dyad which we will define shortly.\footnote{Lower case letters (tensor indices) range from 1-4, capitalised (spinor indices) range from 0-1.}  We first introduce some basic definitions and relations of Riemannian geometry that we find useful~\cite{stephani_exact_2009}.  Let us denote by $\nabla_{a}$ the connection or covariant derivative which defines covariant differentiation.  It satisfies
\begin{equation}
\label{metriccompatibility}
\nabla_ag_{bc} = 0,
\end{equation}
where $g_{ab} = g_{ba}$ is a symmetric non-singular tensor called the metric.  In addition to \eqref{metriccompatibility}, called metric-compatibility, the connection satisfies
\begin{equation}
\label{torsionfree}
(\nabla_{a}\nabla_{b} - \nabla_{b}\nabla_{a} )S = 0,
\end{equation}
where $S$ is an arbitrary scalar.  Equation~\eqref{torsionfree} is equivalent to the vanishing of the torsion tensor and therefore the connection will be torsion-free.  If both \eqref{metriccompatibility} and \eqref{torsionfree} are satisfied the connection is uniquely defined by the metric.  We use the following definition of the Riemann tensor in terms of a metric-compatible and torsion-free covariant derivative
\begin{equation}
\label{covariantderivativetoriemann}
(\nabla_{a}\nabla_{b} - \nabla_{b}\nabla_{a} )V^{d} = {R_{abc}}^{d}V^{c},
\end{equation}
where $V^{a}$ is an arbitrary vector.  With all indices lowered, it has the following symmetries
\begin{equation}
\label{riemannsymmetries}
R_{abcd} = -R_{bacd}, \quad R_{abcd} = -R_{abdc}, \quad R_{abcd} = R_{cdab}, \quad R_{[abc]d}=0,
\end{equation}
where square brackets denote antisymmetrization over the enclosed indices and round brackets denote symmetrization. The Ricci tensor is formed from the contraction of the Riemann tensor
\begin{equation}
\label{Riccitensordefinition}
R_{ab} = {R_{acb}}^{c} = g^{cd}R_{acbd} = R_{ba},
\end{equation}
and the Ricci scalar is formed from the contraction of the Ricci tensor
\begin{equation}
\label{Ricciscalardefinition}
R \equiv {R_{a}}^a = g^{ab}R_{ab}.
\end{equation}
The part of the Riemann tensor that has all the trace parts removed is called the Weyl tensor
\begin{equation}
\label{weyltensordefinition}
{C_{ab}}^{cd} = {R_{ab}}^{cd} - 2{R_{[a}}^{[c}{g_{b]}}^{d]} + \tfrac{1}{3}R{g_{[a}}^{c}{g_{b]}}^{d}.
\end{equation}
It shares all the symmetries of the Riemann tensor and in addition its trace is zero
\begin{equation}
\label{weylzerotrace}
{C_{acb}}^c = 0.
\end{equation}
As well as the metric based approach, we make use of a complementary method called the tetrad formalism.  Often an orthonormal-tetrad is employed, however, the null-tetrad is more closely related to the spinor calculus.  Let us therefore introduce a null-tetrad of vectors~$l^a$,~$n^a$,~$m^a$,~$\conj{m}^a$, written together as follows
\begin{equation}
\label{nulltetraddef}
{e_{\textbf{a}}}^a = 
\begin{pmatrix} 
l^a \\
n^a \\
m^a \\
\conj{m}^a
\end{pmatrix},
\end{equation}
%
%
satisfying
\begin{equation}
\label{normalisednulltetrad}
\eta_{\textbf{ab}} = g_{ab} {e_\textbf{a}}^a {e_\textbf{b}}^b,
\end{equation}
where~$\eta_{\textbf{ab}}$ are the components of the following matrix
\begin{equation}
\label{minkowskimetricnullcoordinates}
\eta_{\textbf{ab}} = 
\begin{pmatrix} 
0 & 1 & 0 & 0 \\
1 & 0 & 0 & 0 \\
0 & 0 & 0 & -1\\
0 & 0 & -1 & 0
\end{pmatrix}.
\end{equation}
The null-tetrad may be used to take components of an arbitrary vector $V^a$ in the following way
\begin{equation}
\label{tetradcomponents}
V^\textbf{a} = {e_a}^{\textbf{a}}V^a.
\end{equation}
\\For the study of manifolds and their metrics, there exists an alternative to the tensor calculus called the spinor calculus~\cite{penrose_spinors_1984}.  Tensors then appear to be a particular type of spinor, specifically a spinor whose indices always occur in pairs, one of which is unprimed and the other primed.  That is, we can correlate an abstract tensor index,~$a$, to a pair of abstract spinor indices,~$AA'$.  For example, the covariant derivative is written in the spinor calculus as
\begin{equation}
	\label{spinornabla}
		\nabla_{a}=\nabla_{AA'}.
\end{equation}
In order to incorporate tensors into the spinor calculus, in addition to the two types of indices, primed and unprimed, we require an operation of complex conjugation.  The general rule for complex conjugation of an arbitrary spinor $\kappa^{AB'}$ is
\begin{equation}
\label{complexconjugation}
\widebar{\kappa^{AB'}} = \conj{\kappa}^{A'B},	
\end{equation}
where the complex conjugate, denoted with an over-bar, is obtained by replacing all unprimed indices with primed indices and vice versa.  In the case of a scalar, \eqref{complexconjugation} is equivalent to the standard complex conjugacy relation.  The condition for the spinor $\kappa^{AB'}$ to be real is
\begin{equation}
\label{realspinor}
\kappa^{AB'} = \conj{\kappa}^{AB'},
\end{equation}
a particular example is the covariant derivative
\begin{equation}
\label{realcovariantderivative}
\nabla_{AB'} = \conj{\nabla}_{AB'}.
\end{equation}
The metric tensor may be legitimately written in terms of abstract indices as
%
\begin{equation}
\label{metricspinor}
g_{ab} = \varepsilon_{AB}{\varepsilon}_{A'B'},
\end{equation}
%
%
where
\begin{equation}
\label{antisymmetricepsilonspinor}
\varepsilon_{AB} = -\varepsilon_{BA} \quad {\varepsilon}_{A'B'} = -{\varepsilon}_{B'A'},
\end{equation}
and where the conjugate spinor, $\varepsilon_{A'B'}$, is obtained from $\varepsilon_{AB}$ by complex conjugation.\footnote{In the case of $\varepsilon_{A'B'}$, the over-bar is usually omitted.}
The component form of \eqref{metricspinor} requires the introduction of the \emph{Infeld-van der Waerden symbols}~\cite{infeld_wellengleichung_1933} and would explicitly represent a change of basis from a tensor basis to a spinor basis.  From \eqref{metricspinor} we see that \eqref{antisymmetricepsilonspinor} will play a similar role to the metric tensor of the tensor calculus, however, there are important differences arising from its antisymmetry.  It is two-dimensional and raises and lowers spinor indices in the following way
\begin{equation}
\label{epsilonspinor}
\begin{array}{l}                   
\varepsilon_{AB}\kappa^A = \kappa_B,\\
\varepsilon^{AB}\kappa_B = \kappa^A,\\ 
\varepsilon^{AB}\varepsilon_{CB} = {\varepsilon_C}^A,
\end{array}
\end{equation}
where~$\kappa^A$ is an arbitrary spinor and~$\varepsilon^{AB} = -\varepsilon^{BA}$.  Analogous relations hold for the conjugate spinor when raising and lowering primed indices.  We note the minus sign in the following \emph{see-saw} property,
\begin{equation}
\label{seesaw}
\kappa_A\sigma^A = - \kappa^A\sigma_A.
\end{equation}
which stands in contrast to the contraction of tensor indices where a plus sign appears instead.
%
%
Analogous to the tetrad in the tensor calculus, in the spinor calculus we have a normalised dyad
\begin{equation}
\label{dyad}
{\varepsilon_{\textbf{A}}}^A = (o^A,\iota^A),
\end{equation}
which satisfies the following relations,\footnote{This is a two-dimensional complex spinor basis, when normalised it is also referred to as a spin-frame.}
\begin{equation}
\label{spinframe}
\begin{array}{l}                   
\varepsilon_{AB}o^Ao^B = o_Ao^A = 0,\\
\varepsilon_{AB}\iota^A\iota^B = \iota_A\iota^A = 0,\\ 
\varepsilon_{AB}o^A\iota^B = o_A\iota^A = 1.
\end{array}
\end{equation}
The dyad may be used to take components of an arbitrary spinor with unprimed indices in the following way
\begin{equation}
\label{dyadcomponents}
\kappa^\textbf{A} = {\varepsilon_A}^{\textbf{A}}\kappa^A.
\end{equation}
Analogous relations to \eqref{dyadcomponents} hold for the conjugate dyad when taking components of a spinor with primed indices.
%
%
%
We can express the null tetrad \eqref{nulltetraddef} in terms of the dyad \eqref{dyad} and its conjugate as follows
\begin{equation}
\label{dyadtotetrad}
\begin{array}{l}                   
l^a = o^{A}o^{A'},\\
n^a = \iota^{A}\iota^{A'},\\
m^a = o^{A}\iota^{A'},\\
\conj{m}^a = \iota^{A}o^{A'}.
\end{array}
\end{equation}
\section{Prime and asterisk operations}
\label{primeasterisk}
The significance of the \emph{primed} symbols, e.g. the primed spin-coefficients occurring in \eqref{basistospincoefficients}, is that under the transformation
\begin{equation}
\label{primetransformation}
o^A \mapsto i \iota^A, \quad \iota^A \mapsto i o^A, \quad o^{A'} \mapsto -i \iota^{A'}, \quad \iota^{A'} \mapsto -i o^{A'},
\end{equation}
which preserves the normalisation condition~\eqref{spinframe}, the primed and unprimed spin-coefficients are interchanged.  From \eqref{dyadtotetrad} we see that under the prime operation the null tetrad transforms as follows
\begin{equation}
\label{tetradprime}
\begin{array}{llll}
l^a \mapsto n^a & n^a \mapsto l^a & m^a \mapsto \conj{m}^a & \conj{m}^a \mapsto m^a.
\end{array}                          
\end{equation}
Furthermore, the curvature spinor components transform as
\begin{equation}
\label{weylricciprime}
\begin{array}{llll}           
\Psi_{r} \mapsto \Psi_{s} & 0 \leftrightarrow 4 & 1 \leftrightarrow 3 & 2 \leftrightarrow 2,\\
\Phi_{rs} \mapsto \Phi_{tu} & 0 \leftrightarrow 2 & 1 \leftrightarrow 1,
\end{array}                          
\end{equation}
and the Bach spinor components transform in the same way as the Ricci spinor components.  The \emph{prime operation} is then defined to be \eqref{primetransformation}.  Note that the operation of complex conjugation, which interchanges primed indices for unprimed indices, commutes with the prime operation such that~$(\conj{\alpha})'=\widebar{(\alpha')}$.  There exists another discrete symmetry possessed by the spin-coefficient formalism called the asterisk (*) operation which effects the following transformation
\begin{equation}
	\label{asterisktransformation}
		o^A \mapsto o^A, \quad \iota^A \mapsto  \iota^A, \quad o^{A'} \mapsto \iota^{A'}, \quad \iota^{A'} \mapsto -o^{A'}.
\end{equation}
From \eqref{asterisktransformation} and the definitions \eqref{dyadtotetrad}, \eqref{basistospincoefficients}, \eqref{intrinsicderivative}, \eqref{spinorcomponents} the corresponding transformation under the asterisk operation of the null tetrad, spin-coefficients, intrinsic derivatives and spinor components follow respectively.
The prime and asterisk operations may be used to generate new equations from equations already known, or alternatively to check the correctness and consistency of equations found by other means.
\section{Newman-Penrose equations}
\label{newmanPenrose}
Equation \eqref{covariantderivativetoriemann} defines the Riemann tensor in terms of the commutator of the covariant derivatives.  In the spinor formalism there is an analogous relation when such commutators are applied to spinors
\begin{align}
\label{spinorcovariantderivativetoriemann}
&({\varepsilon}_{A'B'}\nabla_{C'(A}{\nabla_{B)}}^{C'} + \varepsilon_{AB}\nabla_{C(A'}{\nabla_{B')}}^{C})\kappa^{D}\nonumber\\ &= (\varepsilon_{A'B'}{\Psi_{ABD}}^{C} + \Lambda{\varepsilon}_{A'B'}(\varepsilon_{AD}{\varepsilon_{B}}^{C} + {\varepsilon_{A}}^{C}\varepsilon_{BD}) + \varepsilon_{AB}{\Phi_{A'B'D}}^C)\kappa^{D},
%
\end{align}
%
where~$\kappa^{D}$ is an arbitrary spinor.  Setting $\kappa^{D} = {\varepsilon_{\textbf{D}}}^{D}$ and taking components of \eqref{spinorcovariantderivativetoriemann}, we can use \eqref{intrinsicderivativetocompactedintrinsicderivative} to obtain the \emph{Newman-Penrose} equations in the GHP formalism~\cite{penrose_spinors_1984}
\begin{subequations}
	\label{compactednpequations}
	\begin{align}
	\tho\rho - \edh'\kappa & = \rho^2 + \sigma\conj{\sigma} - \conj{\kappa}\tau - \tau'\kappa + \Phi_{00}, \\
	%
	%
	\tho\sigma - \edh\kappa & = (\rho + \conj{\rho})\sigma - (\tau + \conj{\tau}')\kappa + \Psi_0, \\
	%
	%
	\tho\tau - \tho'\kappa & = (\tau - \conj{\tau}')\rho + (\conj{\tau} - \tau')\sigma + \Psi_1 + \Phi_{01}, \\
	%
	%
	\edh\rho - \edh'\sigma & = (\rho - \conj{\rho})\tau + (\conj{\rho}' - \rho')\kappa - \Psi_1 + \Phi_{01}, \\
	%
	%
	\edh\tau - \tho'\sigma & = -\rho'\sigma - \conj{\sigma}'\rho + \tau^2 + \kappa\conj{\kappa}' + \Phi_{02}, \\
	%
	%
	\tho'\rho - \edh'\tau & = \rho\conj{\rho}' + \sigma\sigma' - \tau\conj{\tau} - \kappa\kappa' - \Psi_2 - 2\Lambda.
	%
	\end{align}
\end{subequations}
Applying the prime operation to \eqref{compactednpequations} we obtain six more relations.  Equations \eqref{compactednpequations} and their primed partners are equivalent to the Ricci identities of the tensor calculus~\cite{stephani_exact_2009}.
The remaining Newman-Penrose equations take the form of commutator expressions.  To derive them, consider the following commutator of intrinsic derivatives acting on an arbitrary scalar $f$,
\begin{equation}
\label{intrinsicderivativecommutator}
\left[\nabla_{\textbf{AB}'}, \nabla_{\textbf{CD}'}\right]f = \left(\nabla_{\textbf{AB}'}({\varepsilon_{\textbf{C}}}^{Q}{{\varepsilon}_{\textbf{D}'}}^{Q'}) - \nabla_{\textbf{CD}'}({\varepsilon_{\textbf{A}}}^{Q}{{\varepsilon}_{\textbf{B}'}}^{Q'})\right)\nabla_{QQ'}f.
\end{equation}
Taking components of \eqref{intrinsicderivativecommutator} and using \eqref{intrinsicderivativetocompactedintrinsicderivative}, the following commutator expressions can be derived
\begin{subequations}
	\label{compactednpcommutators}
	\begin{align}
	\tho\tho' - \tho'\tho = {}&(\conj{\tau} - \tau')\edh + (\tau - \conj{\tau}')\edh' - p(\kappa\kappa' - \tau\tau' + \Psi_2 + \Phi_{11} - \Lambda) \nonumber \\ &- q(\conj{\kappa}\conj{\kappa}' - \conj{\tau}\conj{\tau}' + \conj{\Psi}_2 + \Phi_{11} - \Lambda), \label{thotho'} \\
	\edh\edh' - \edh'\edh = {}&(\conj{\rho}' - \rho')\tho + (\rho - \conj{\rho})\tho' + p(\rho\rho' - \sigma\sigma' + \Psi_2 - \Phi_{11} - \Lambda) \nonumber \\ &- q(\conj{\rho}\conj{\rho}' - \conj{\sigma}\conj{\sigma}' + \conj{\Psi}_2 - \Phi_{11} - \Lambda), \label{etheth'}	\\
	\tho\edh - \edh\tho = {}&\conj{\rho}\edh +\sigma\edh' - \conj{\tau}'\tho - \kappa\tho' - p(\rho'\kappa - \tau'\sigma + \Psi_1) - q(\conj{\sigma}'\conj{\kappa} - \conj{\rho}\conj{\tau}' + \Phi_{01}). \label{thoeth}
	%
	%
	%
	\end{align}
\end{subequations}
Three more commutator equations are obtained by priming, conjugating and both priming and conjugating \eqref{thoeth}.
%
\section{Spinor form of the Bianchi identity}
\label{bianchiidentity}
The Riemann tensor satisfies an important differential identity called the \emph{Bianchi identity}
\begin{equation}
\label{tensorbianchiidentity}
	\nabla_{[a}R_{bc]de} = 0.
\end{equation}
\eqref{tensorbianchiidentity} is equivalent to~\cite{lanczos_remarkable_1938}
\begin{equation}
\label{divergencetensorbianchiidentity}
	\nabla^{a}\tensor[^*]{R}{_{abcd}} = 0,
\end{equation}
where we define the (left) Hodge dual operation to act on the first and second indices as follows 
\begin{equation}
\label{lefthodgedualdefinition}
	\tensor[^*]{R}{_{abcd}} = \tfrac{1}{2}{e_{ab}}^{pq}R_{pqcd},
\end{equation}
and where $e_{abcd} = e_{[abcd]}$ is the alternating tensor.\footnote{Also known as the four-dimensional Levi-Civita tensor.}
Substituting \eqref{weyltensordefinition} into \eqref{divergencetensorbianchiidentity} and making use of \eqref{lefthodgedualdefinition}
%
%
%
gives the following differential relation between the Weyl tensor, Ricci tensor and Ricci scalar
\begin{equation}
\label{weylriccibianchiidentity}
\nabla^{a}C_{abcd} = \nabla_{[c}R_{d]b} + \tfrac{1}{6}g_{b[c}\nabla_{d]}R.
\end{equation}
Taking the trace of \eqref{weylriccibianchiidentity} gives
\begin{equation}
\label{tracebianchiidentity}
\nabla^{a}R_{ab}-\tfrac{1}{2}\nabla_{b}R = 0.
\end{equation}
The spinor equivalent of \eqref{weylriccibianchiidentity} is
\begin{equation}
\label{bianchiidentityspinor}
\nabla^{A}_{B'}\Psi_{ABCD} = \nabla^{A'}_{B}\Phi_{CDA'B'} - 2\varepsilon_{B(C}\nabla_{D)B'}\Lambda.
\end{equation}
%
%
In terms of the GHP formalism, the components of \eqref{bianchiidentityspinor} are given by the following six equations
\begin{subequations}
	\label{compactedbianchiidentityspinor}
	\begin{align}
	\tho\Psi_1 & - \edh'\Psi_0 - \tho\Phi_{01} + \edh\Phi_{00}\nonumber\\ &
	= -\tau'\Psi_0 + 4\rho\Psi_1 - 3\kappa\Psi_2 + \conj{\tau}'\Phi_{00} - 2\conj{\rho}\Phi_{01} - 2\sigma\Phi_{10} + 2\kappa\Phi_{11} + \conj{\kappa}\Phi_{02},\label{compactedbianchiidentityspinor1}\\
	%
	%
	\tho\Psi_2 & - \edh'\Psi_1 - \edh'\Phi_{01} + \tho'\Phi_{00} + 2\tho\Lambda\nonumber\\ &
	= \sigma'\Psi_0 - 2\tau'\Psi_1 + 3\rho\Psi_2 - 2\kappa\Psi_3 + 2\conj{\rho}'\Phi_{00} - 2\conj{\tau}\Phi_{01} - 2\tau\Phi_{10} + 2\rho\Phi_{11} + \conj{\sigma}\Phi_{02},\label{compactedbianchiidentityspinor3}\\
	%
	%
	\tho\Psi_3 & - \edh'\Psi_2 - \tho\Phi_{21} + \edh\Phi_{20} - 2\edh'\Lambda\nonumber\\ &
	= 2\sigma'\Psi_1 - 3\tau'\Psi_2 + 2\rho\Psi_3 - \kappa\Psi_4 - 2\rho'\Phi_{10} + 2\tau'\Phi_{11} + \conj{\tau}'\Phi_{20} - 2\conj{\rho}\Phi_{21} + \conj{\kappa}\Phi_{22},\label{compactedbianchiidentityspinor5}\\
	%
	%
	\tho\Psi_4 & - \edh'\Psi_3 - \edh'\Phi_{21} + \tho'\Phi_{20}\nonumber\\ &
	= 3\sigma'\Psi_2 - 4\tau'\Psi_3 + \rho\Psi_4 - 2\kappa'\Phi_{10} + 2\sigma'\Phi_{11} + \conj{\rho}'\Phi_{20} - 2\conj{\tau}\Phi_{21} + \conj{\sigma}\Phi_{22},\label{compactedbianchiidentityspinor7}\\
	%
	%
	\tho\Phi_{12} & + \tho'\Phi_{01} - \edh\Phi_{11} - \edh'\Phi_{02} + 3\edh\Lambda\nonumber\\
	&= \left(\rho' + 2\conj{\rho}'\right)\Phi_{01} + \left(2\rho + \conj{\rho}\right)\Phi_{12} - \left(\tau' + \conj{\tau}\right)\Phi_{02} - 2\left(\tau + \conj{\tau}'\right)\Phi_{11}\nonumber\\ 
	&- \conj{\kappa}'\Phi_{00} - \kappa\Phi_{22} + \sigma\Phi_{21} + \conj{\sigma}'\Phi_{10},\label{compactedbianchiidentityspinor9}\\
	%
	%
	\tho\Phi_{11} & + \tho'\Phi_{00} - \edh\Phi_{10} - \edh'\Phi_{01} + 3\tho\Lambda\nonumber\\
	&= \left(\rho' + \conj{\rho}'\right)\Phi_{00} + 2\left(\rho + \conj{\rho}\right)\Phi_{11} - \left(\tau' + 2\conj{\tau}\right)\Phi_{01} - \left(2\tau + \conj{\tau}'\right)\Phi_{10}\nonumber\\ 
	&- \conj{\kappa}\Phi_{12} - \kappa\Phi_{21} + \sigma\Phi_{20} + \conj{\sigma}\Phi_{02},\label{compactedbianchiidentityspinor11}
	%
	\end{align}
\end{subequations}
and their corresponding primed relations.
%
%
%
\section{Calculating spin-coefficients using Cartan's calculus}
\label{cartancalculus}
The spin-coefficients can be calculated efficiently using Cartan's differential calculus~\cite{penrose_spinors_1984}.  First we take the exterior derivative of the dual basis one-forms,
\begin{align}
\label{tetradtospincoefficients}
\dd\textbf{l} &=l_{[\textbf{b},\textbf{a}]}e^{\textbf{ma}}e^{\textbf{nb}}(\textbf{e}_\textbf{m} \wedge \textbf{e}_\textbf{n}), \\
\dd\textbf{n} &=n_{[\textbf{b},\textbf{a}]}e^{\textbf{ma}}e^{\textbf{nb}}(\textbf{e}_\textbf{m} \wedge \textbf{e}_\textbf{n}), \\
\dd\textbf{m} &=m_{[\textbf{b},\textbf{a}]}e^{\textbf{ma}}e^{\textbf{nb}}(\textbf{e}_\textbf{m} \wedge \textbf{e}_\textbf{n}),
\end{align}
where a comma denotes partial differentiation and
\begin{equation}
\label{dualbasis}
\textbf{e}_1 = \textbf{l}, \textbf{e}_2 = \textbf{n}, \textbf{e}_3 = \textbf{m}, \textbf{e}_4 = \conj{\textbf{m}},
\end{equation}
which gives
%
%
\begin{subequations}
	\label{exteriorderivativeofnulltetrad}
	\begin{align}
	\dd\textbf{l} = {}&l_{[\textbf{b},\textbf{a}]}\left(e^{1\textbf{a}}e^{2\textbf{b}}(\textbf{l} \wedge \textbf{n}) + e^{1\textbf{a}}e^{3\textbf{b}}(\textbf{l} \wedge \textbf{m}) + e^{1\textbf{a}}e^{4\textbf{b}}(\textbf{l} \wedge \conj{\textbf{m}})\right) \nonumber \\ & + l_{[\textbf{b},\textbf{a}]}\left(e^{2\textbf{a}}e^{3\textbf{b}}(\textbf{n} \wedge \textbf{m}) + e^{2\textbf{a}}e^{4\textbf{b}}(\textbf{n} \wedge \conj{\textbf{m}}) + e^{3\textbf{a}}e^{4\textbf{b}}(\textbf{m} \wedge \conj{\textbf{m}})\right), \\ \nonumber
	\dd\textbf{n}  = {}&n_{[\textbf{b},\textbf{a}]}\left(e^{1\textbf{a}}e^{2\textbf{b}}(\textbf{l} \wedge \textbf{n}) + e^{1\textbf{a}}e^{3\textbf{b}}(\textbf{l} \wedge \textbf{m}) + e^{1\textbf{a}}e^{4\textbf{b}}(\textbf{l} \wedge \conj{\textbf{m}})\right) \nonumber \\ & + n_{[\textbf{b},\textbf{a}]}\left(e^{2\textbf{a}}e^{3\textbf{b}}(\textbf{n} \wedge \textbf{m}) + e^{2\textbf{a}}e^{4\textbf{b}}(\textbf{n} \wedge \conj{\textbf{m}}) + e^{3\textbf{a}}e^{4\textbf{b}}(\textbf{m} \wedge \conj{\textbf{m}})\right), \\ \nonumber
	\dd\textbf{m}  = {}&m_{[\textbf{b},\textbf{a}]}\left(e^{1\textbf{a}}e^{2\textbf{b}}(\textbf{l} \wedge \textbf{n}) + e^{1\textbf{a}}e^{3\textbf{b}}(\textbf{l} \wedge \textbf{m}) + e^{1\textbf{a}}e^{4\textbf{b}}(\textbf{l} \wedge \conj{\textbf{m}})\right) \nonumber \\ & + m_{[\textbf{b},\textbf{a}]}\left(e^{2\textbf{a}}e^{3\textbf{b}}(\textbf{n} \wedge \textbf{m}) + e^{2\textbf{a}}e^{4\textbf{b}}(\textbf{n} \wedge \conj{\textbf{m}}) + e^{3\textbf{a}}e^{4\textbf{b}}(\textbf{m} \wedge \conj{\textbf{m}})\right).
	\end{align}
\end{subequations}
The spin-coefficients are defined as follows,
\begin{subequations}
	\label{exteriorderivativeofnulltetradtospincoefficents}
	\begin{align}
	2\dd\textbf{l} = {}&(\gamma' + \conj{\gamma}')(\textbf{l} \wedge \textbf{n}) + (\conj{\beta} - \beta'- \conj{\tau})(\textbf{l} \wedge \textbf{m}) + (\beta - \conj{\beta}' - \tau)(\textbf{l} \wedge \conj{\textbf{m}}) \nonumber \\ & - \conj{\kappa}(\textbf{n} \wedge \textbf{m}) - \kappa(\textbf{n} \wedge \conj{\textbf{m}}) + (\rho - \conj{\rho})(\textbf{m} \wedge \conj{\textbf{m}}),\label{exteriorofl} \\	\nonumber
	2\dd\textbf{n} = {}&-(\gamma + \bar{\gamma})(\textbf{l} \wedge \textbf{n}) - \kappa'(\textbf{l} \wedge \textbf{m}) - \conj{\kappa}'(\textbf{l} \wedge \conj{\textbf{m}}) \nonumber \\ & + (-\tau' + \beta' - \conj{\beta})(\textbf{n} \wedge \textbf{m}) + (-\conj{\tau}' + \conj{\beta}' - \beta)(\textbf{n} \wedge \conj{\textbf{m}}) - (\rho' - \conj{\rho}')(\textbf{m} \wedge \conj{\textbf{m}}),\label{exteriorofn} \\ \nonumber
	2\dd\textbf{m} = {}&-(\tau - \conj{\tau}')(\textbf{l} \wedge \textbf{n}) + (\gamma - \conj{\gamma} - \conj{\rho}')(\textbf{l} \wedge \textbf{m}) - \conj{\sigma}'(\textbf{l} \wedge \conj{\textbf{m}}) \nonumber \\ & + (-\gamma' + \conj{\gamma}' - \rho)(\textbf{n} \wedge \textbf{m}) - \sigma(\textbf{n} \wedge \conj{\textbf{m}}) + (\beta + \conj{\beta}')(\textbf{m} \wedge \conj{\textbf{m}}).\label{exteriorofm}
	\end{align}
\end{subequations}
Comparing \eqref{exteriorderivativeofnulltetradtospincoefficents} with \eqref{exteriorderivativeofnulltetrad} we obtain
\begin{equation}
\label{solutionlnmbasisvectors}
\begin{array}{lll}
\gamma' + \conj{\gamma}' = 2l_{[\textbf{b},\textbf{a}]}n^\textbf{a}l^\textbf{b}, & -\conj{\beta} + \beta' + \conj{\tau} = 2l_{[\textbf{b},\textbf{a}]}l^\textbf{a}\conj{m}^\textbf{b}, & -\beta + \conj{\beta}' + \tau = 2l_{[\textbf{b},\textbf{a}]}n^\textbf{a}m^\textbf{b},\\
\conj{\kappa} = 2l_{[\textbf{b},\textbf{a}]}l^\textbf{a}\conj{m}^\textbf{b}, & \kappa = 2l_{[\textbf{b},\textbf{a}]}l^\textbf{a}m^\textbf{b}, & \rho - \conj{\rho} = 2l_{[\textbf{b},\textbf{a}]}\conj{m}^\textbf{a}m^\textbf{b},\\
-\gamma - \bar{\gamma} = 2n_{[\textbf{b},\textbf{a}]}n^\textbf{a}l^\textbf{b}, & \kappa' = 2n_{[\textbf{b},\textbf{a}]}n^\textbf{a}\conj{m}^\textbf{b}, & \conj{\kappa}' = 2n_{[\textbf{b},\textbf{a}]}n^\textbf{a}m^\textbf{b},\\
\tau' - \beta' + \conj{\beta} = 2n_{[\textbf{b},\textbf{a}]}l^\textbf{a}\conj{m}^\textbf{b}, & \conj{\tau}' - \conj{\beta}' + \beta = 2n_{[\textbf{b},\textbf{a}]}l^\textbf{a}m^\textbf{b}, & -\rho' + \conj{\rho}' = 2n_{[\textbf{b},\textbf{a}]}\conj{m}^\textbf{a}m^\textbf{b},\\
-\tau + \conj{\tau}' = 2m_{[\textbf{b},\textbf{a}]}n^\textbf{a}l^\textbf{b}, &-\gamma + \conj{\gamma} + \conj{\rho}' = 2m_{[\textbf{b},\textbf{a}]}n^\textbf{a}\conj{m}^\textbf{b}, &  \conj{\sigma}' = 2m_{[\textbf{b},\textbf{a}]}n^\textbf{a}m^\textbf{b},\\
\gamma' - \conj{\gamma}' + \rho = 2m_{[\textbf{b},\textbf{a}]}l^\textbf{a}\conj{m}^\textbf{b}, & \sigma = 2m_{[\textbf{b},\textbf{a}]}l^\textbf{a}m^\textbf{b}, & \beta + \conj{\beta}' = 2m_{[\textbf{b},\textbf{a}]}\conj{m}^\textbf{a}m^\textbf{b}.
\end{array}
\end{equation}
Equations \eqref{solutionlnmbasisvectors} actually contain only 24 independent real equations, and thus enable us to solve uniquely for the twelve complex spin-coefficients in terms of the null tetrad~\cite{cocke_table_1989}.
\section*{Acknowledgements}
The author acknowledges with thanks the advice and encouragement of Prof. A. Fring and the doctoral studentship provided by City, University of London.
%
\bibliography{The_Bach_equations_in_Spin_Coefficient_form}      		 

\providecommand{\href}[2]{#2}\begingroup\raggedright\begin{thebibliography}{10}

\bibitem{capozziello_extended_2011}
S.~Capozziello and M.~De~Laurentis, ``Extended {Theories} of {Gravity},''
  \href{http://dx.doi.org/10.1016/j.physrep.2011.09.003}{{\em Physics Reports}
  {\bfseries 509} no.~4, (Dec., 2011) 167--321}.
  \url{http://www.sciencedirect.com/science/article/pii/S0370157311002432}.

\bibitem{schmidt_fourth_2007}
H.-J. Schmidt, ``Fourth order gravity: equations, history, and applications to
  cosmology,'' \href{http://dx.doi.org/10.1142/S0219887807001977}{{\em
  International Journal of Geometric Methods in Modern Physics} {\bfseries 04}
  no.~02, (Mar., 2007) 209--248}.
  \url{https://www.worldscientific.com/doi/abs/10.1142/S0219887807001977}.

\bibitem{bach_zur_1921}
R.~Bach, ``Zur {Weylschen} {Relativit{\"a}tstheorie} und der {Weylschen}
  {Erweiterung} des {Kr{\"u}mmungstensorbegriffs},''
  \href{http://dx.doi.org/10.1007/BF01378338}{{\em Mathematische Zeitschrift}
  {\bfseries 9} no.~1, (Mar., 1921) 110--135}.
  \url{https://doi.org/10.1007/BF01378338}.

\bibitem{fiedler_exact_1980}
B.~Fiedler and R.~Schimming, ``Exact solutions of the {Bach} field equations of
  general relativity,'' {\em Reports on Mathematical Physics} {\bfseries 17}
  no.~1, (1980) 15--36.

\bibitem{mannheim_exact_1989}
P.~D. Mannheim and D.~Kazanas, ``Exact vacuum solution to conformal {Weyl}
  gravity and galactic rotation curves,'' {\em The Astrophysical Journal}
  {\bfseries 342} (1989) 635--638.

\bibitem{mannheim_solutions_1991}
P.~D. Mannheim and D.~Kazanas, ``Solutions to the {Reissner}-{Nordstr{\"o}m},
  {Kerr}, and {Kerr}-{Newman} problems in fourth-order conformal {Weyl}
  gravity,'' \href{http://dx.doi.org/10.1103/PhysRevD.44.417}{{\em Physical
  Review D} {\bfseries 44} no.~2, (July, 1991) 417--423}.
  \url{https://link.aps.org/doi/10.1103/PhysRevD.44.417}.

\bibitem{newman_approach_1962}
E.~Newman and R.~Penrose, ``An approach to gravitational radiation by a method
  of spin coefficients,'' {\em Journal of Mathematical Physics} {\bfseries 3}
  no.~3, (1962) 566--578.

\bibitem{geroch_space-time_1973}
R.~Geroch, A.~Held, and R.~Penrose, ``A space-time calculus based on pairs of
  null directions,'' {\em Journal of Mathematical Physics} {\bfseries 14}
  no.~7, (1973) 874--881.

\bibitem{stephani_exact_2009}
H.~Stephani, D.~Kramer, M.~MacCallum, C.~Hoenselaers, and E.~Herlt, {\em Exact
  solutions of {Einstein}'s field equations}.
\newblock Cambridge University Press, 2009.

\bibitem{weyl_gravitation_1918}
H.~Weyl, ``Gravitation und {Elektrizit{\"a}t},'' {\em Sitzungsber. K{\"o}nigl.
  Preuss. Akad. Wiss.} {\bfseries 26} (1918) 465--80.

\bibitem{lanczos_electricity_1957}
C.~Lanczos, ``Electricity and {General} {Relativity},''
  \href{http://dx.doi.org/10.1103/RevModPhys.29.337}{{\em Reviews of Modern
  Physics} {\bfseries 29} no.~3, (July, 1957) 337--350}.
  \url{https://link.aps.org/doi/10.1103/RevModPhys.29.337}.

\bibitem{t_hooft_one_1974}
G.~'t~Hooft and M.~J.~G. Veltman, ``One loop divergencies in the theory of
  gravitation,'' {\em Ann.Inst.H.Poincare Phys.Theor.} {\bfseries A20} (1974)
  69--94.

\bibitem{stelle_renormalization_1977}
K.~S. Stelle, ``Renormalization of higher-derivative quantum gravity,''
  \href{http://dx.doi.org/10.1103/PhysRevD.16.953}{{\em Physical Review D}
  {\bfseries 16} no.~4, (Aug., 1977) 953--969}.
  \url{https://link.aps.org/doi/10.1103/PhysRevD.16.953}.

\bibitem{stelle_classical_1978}
K.~S. Stelle, ``Classical gravity with higher derivatives,''
  \href{http://dx.doi.org/10.1007/BF00760427}{{\em General Relativity and
  Gravitation} {\bfseries 9} no.~4, (Apr., 1978) 353--371}.
  \url{https://link.springer.com/article/10.1007/BF00760427}.

\bibitem{penrose_spinors_1984}
R.~Penrose and W.~Rindler, {\em Spinors and {Space}-{Time}: {Volume} 1,
  {Two}-{Spinor} {Calculus} and {Relativistic} {Fields}}.
\newblock Cambridge University Press, 1984.

\bibitem{ehlers_exact_1962}
J.~Ehlers and W.~Kundt, ``Exact solutions of the gravitational field
  equations,'' in {\em The {Theory} of {Gravitation}}, L.~Witten, ed.,
  pp.~49--101.
\newblock John Wiley \& Sons, Inc., 1962.

\bibitem{madsen_plane_1990}
M.~S. Madsen, ``The plane gravitational wave in quadratic gravity,'' {\em
  Classical and Quantum Gravity} {\bfseries 7} no.~1, (1990) 87.

\bibitem{kozameh_conformal_1985}
C.~N. Kozameh, E.~T. Newman, and K.~Tod, ``Conformal {Einstein} spaces,'' {\em
  General relativity and gravitation} {\bfseries 17} no.~4, (1985) 343--352.

\bibitem{szekeres_conformal_1968}
P.~Szekeres, ``Conformal tensors,''
  \href{http://dx.doi.org/10.1098/rspa.1968.0076}{{\em Proc. R. Soc. Lond. A}
  {\bfseries 304} no.~1476, (Apr., 1968) 113--122}.
  \url{http://rspa.royalsocietypublishing.org/content/304/1476/113}.

\bibitem{brinkmann_einstein_1925}
H.~W. Brinkmann, ``Einstein spaces which are mapped conformally on each
  other,'' \href{http://dx.doi.org/10.1007/BF01208647}{{\em Mathematische
  Annalen} {\bfseries 94} no.~1, (Dec., 1925) 119--145}.
  \url{https://link.springer.com/article/10.1007/BF01208647}.

\bibitem{penrose_remarkable_1965}
R.~Penrose, ``A {Remarkable} {Property} of {Plane} {Waves} in {General}
  {Relativity},'' \href{http://dx.doi.org/10.1103/RevModPhys.37.215}{{\em
  Reviews of Modern Physics} {\bfseries 37} no.~1, (Jan., 1965) 215--220}.
  \url{https://link.aps.org/doi/10.1103/RevModPhys.37.215}.

\bibitem{davis_simple_1976-1}
T.~M. Davis, ``A simple application of the {Newman}-{Penrose} spin coefficient
  formalism. {I},'' {\em International Journal of Theoretical Physics}
  {\bfseries 15} no.~5, (1976) 315--317.

\bibitem{chandrasekhar_mathematical_1998}
S.~Chandrasekhar, {\em The mathematical theory of black holes}, vol.~69.
\newblock Oxford university press, 1998.

\bibitem{infeld_wellengleichung_1933}
L.~Infeld and B.~L. van~der Waerden, ``Die {Wellengleichung} des {Elektrons} in
  der allgemeinen {Relativit{\"a}tstheorie},'' {\em Sitzungsber. Ber. Preuss.
  Akad. Wiss. Physik.-math.} {\bfseries 9} (1933) 380--401.

\bibitem{lanczos_remarkable_1938}
C.~Lanczos, ``A remarkable property of the {Riemann}-{Christoffel} tensor in
  four dimensions,'' {\em Annals of Mathematics} (1938) 842--850.

\bibitem{cocke_table_1989}
W.~J. Cocke, ``Table for constructing the spin coefficients in general
  relativity,'' \href{http://dx.doi.org/10.1103/PhysRevD.40.650}{{\em Physical
  Review D} {\bfseries 40} no.~2, (July, 1989) 650--651}.
  \url{https://link.aps.org/doi/10.1103/PhysRevD.40.650}.

\end{thebibliography}\endgroup
\end{document}